\documentclass[12pt,a4paper,english]{article}
\usepackage[latin1]{inputenc}
\usepackage{graphicx}
\begin{document}

\title{Are scale-free regulatory networks larger than random ones?}
\author{Miguel A. Fortuna{*} and Carlos J. Meli\'an}
\date{Integrative Ecology Group \\
 Estación Biológica de Doñana, CSIC \\
 Apdo. 1056, E-41080 Sevilla, Spain\\}

\maketitle

\begin{center}
{\sc Running Title}:\\ \large{Scale-free Regulatory Networks}
\end{center}

\vfill
{*}Correspondence should be addressed to:
fortuna@ebd.csic.es\\
Phone: +34 954-23-23-40; 
Fax: +34 954-62-11-25

\newpage
\begin{abstract}
\normalsize{Network of packages with regulatory interactions (dependences 
and conflicts) from Debian GNU/Linux operating system is compiled and used 
as analogy of a gene regulatory network.
Using a trace-back algorithm we assembly networks from the potential 
pool of packages for both scale-free and exponential topology from real  
and a null model data, respectively.
We calculate the maximum number of packages that can be 
functionally installed in the system (i.e., the active network size).
We show that scale-free regulatory networks allow a larger active network 
size than random ones.
Small genomes with scale-free regulatory topology could allow much more
functionality than large genomes with an exponential one, with implications
on its dynamics, robustness and evolution.}
\end{abstract}

\newpage

In the last years an increasing number of complex systems have been described
as networks. They are represented by nodes connected between them with different
topological properties, examples of which include biological, technological and
social systems [1-3]. 
Some of them are assembled by several types of interactions [4-6]. 
In particular, regulatory interactions (when nodes can be up or down regulated) at 
genomic scale (in which genes can affect each other's expression) are 
becoming increasingly resolved [7,8].
 
Recent evidences from whole-genome sequence suggest that 
organismal complexity arises much more from elaborate regulation of gene expression 
than by genome size itself [9,10]. 
Previous results on small subsets of genes have shown the importance 
of the topology and the signature of regulatory interactions (i.e., when nodes 
are activated or inhibited) for the robustness of the network [11]. 
But the effects of the topology of regulatory interactions on gene expression 
in large networks are difficult to asses because the small subset of genes 
with known signature of the interactions [7,8,11,12]. 
Could small genomes with scale-free regulatory topology show much more
functionality than large genomes with an exponential one?
Specifically the following question will be addressed here: how  does topology
of regulatory interactions alters the maximum number of activated genes in a 
large regulatory network? 

Despite of genes have associated values that represent concentrations
or levels of activation depending on the values of other cellular units [13],
we can study gene networks through a boolean approach [14].
This only implies the knowledge of the interactions together with their signatures.
Because of this simplification is enough to reproduce the main characteristics
of the regulatory network dynamics [11], we use as analogy of a large gene network 
the most resolved complex network to date with the signature of regulatory interactions. 
Specifically, we have compiled the network of packages 
of Debian GNU/Linux operating system with dependences (activating 
interactions) and conflicts (inhibiting interactions). 

Debian packages network described here is composed by the binary i386 packages
belonging to the sections main, contrib and non-free of the latest
stable Debian distribution (3.0, alias \emph{Woody}), available from
the US Debian Server (http://packages.debian.org/stable)[15]. 
It includes 8,996 nodes (packages), and 
31,904 regulatory interactions (30,003 dependences and 1,901 conflicts). 
Dependence means that package B has to be installed to A 
works, and conflict means that package A does not work if B is installed
in the system.

To test the effect of the topology of a large regulatory network on its 
active network size (the maximum number of activating nodes) we develop a 
null model that (1) preserves the total number of dependences and conflicts 
as in the real network, and (2) maintains statistically the frequency of packages 
with different combinations of incoming and outgoing interactions for dependences 
and conflicts (Fig. 1), forcing them to an exponential degree distribution (Fig. 2a,c). 

We assembled 1,000 replicates from both real data (power law degree distribution) 
and the null model (exponential degree distribution, see Fig. 2a,c) using a 
trace-back algorithm, and counted the total number of packages installed (activated genes)
in each replicate. 
Trace-back algorithm selects randomly a package, checks dependences 
and conflicts of this package with the rest of packages of the network, and 
whether they are installed or not in the network. 
If the package has a conflict with an already installed one, it is discarded (inhibited genes) 
and never will be part of the network. 
If there are no conflicts with installed packages, the algorithm checks whether 
some of the packages on which it depends directly or indirectly (by successive 
dependences), has been discarded or has a conflict with an already
installed package. 
If so, is discarded too. 
Otherwise, is installed with all packages on which it depends directly as 
well as indirectly.
It continues until no more packages are available to be included (i.e., 
packages excluded by the assembly temporal sequence due to their conflicts 
with packages already installed).  
Before starting each replicate, we have automatically installed the 100 packages 
considered basic to the system works [15].
The total number of installed packages represents the active network size of 
each replicate. 
Therefore, as a function of the assembly temporal sequence, each replicate from real 
data and data from the null model has a different number of packages installed.
In this way we obtain the frequency distribution of the active network 
size from both real data and data from the null model (Fig. 2b). 

The frequency distribution of the active network size from data of the null 
model is significantly smaller than from the real data (Fig. 2b). 
Dramatical changes in the active size of complex networks as a function of the topology 
of regulatory interactions can imply differential responses in the robustness 
and functionality of the network [11]. 
Rewiring connections instead of increasing the number of genes seems to be an 
alternative mechanism to enhance the activity of the network [6,8,9]. 
Small genomes with scale-free regulatory topology could 
allow a higher active size than large genomes with an exponential one. 
The present study offers a framework to explore the real ratios of 
activating and inhibiting interactions in large gene networks when data becomes 
available. 
Further work will determine the evolution of active size thresholds 
in scale-free networks when the ratio of both interactions changes.
\\
\\
This work was funded by the Spanish Ministry of Science and Education (Grants
BES-2004-6682 to M.A.F. and FP-2000-6137 to C.J.M.).

\section*{References}
[1] S.H. Strogatz, Nature \textbf{410}, 268 (2001).

[2] R. Albert and A.-L. Barabási, Rev. Mod. Phys. \textbf{74}, 47 (2002).

[3] M.E.J. Newman, SIAM Rev. \textbf{45}, 167 (2003).

[4] R. Axelrod and D. Dion, Science \textbf{242}, 1385-1390 (1988).

[5] E.L. Berlow, A.M. Neutel, J.E. Cohen, and P.C. De Ruiter, J. Anim. Ecol. 
\textbf{73}, 585-598 (2004).

[6] J.M. Stuart, E. Segal, D. Koller, and S.K. Kim, Science \textbf{302}, 
249-255 (2003).

[7] L. Giot, J.S. Bader, C. Brouwer, and A. Chaudhuri, Science \textbf{302}, 
1727-1736 (2003).

[8] N.M. Luscombe, M. Madan Babu, H. Yu, M. Snyder, S. Teichmann, and M. Gerstein,
Nature \textbf{431}, 308-312 (2004).

[9] Knight, J. Nature \textbf{417}, 374-376 (2002).

[10] M. Levine, and R. Tjian, Nature \textbf{424}, 147-151 (2003).

[11] R. Albert, and H.G. Othmer, J. Theor. Biol. \textbf{223}, 1-18 (2003).

[12] E.H. Davidson, J.P. Rast, P. Oliveri, A. Ransick, \emph{et al}., Science 
\textbf{295}, 1669-1678 (2002). 

[13] E.H. Davidson, {\em Genomic Regulatory Systems. Development and 
Evolution}, (Academic Press, San Diego, CA, 2001).

[14] S.A. Kauffman, J. Theor. Biol. \textbf{22}, 437-467 (1969).

[15] Network description files are available from the authors upon request, 
and they contain: 1) code and name of packages (packages.txt), 2) base packages 
(base.txt), and 3) dependences (1) and conflicts (0) between each pair of packages 
(interactions.txt). 

\newpage
\section*{Figures}
Fig-1.
\begin{center}
\includegraphics[width=10cm]{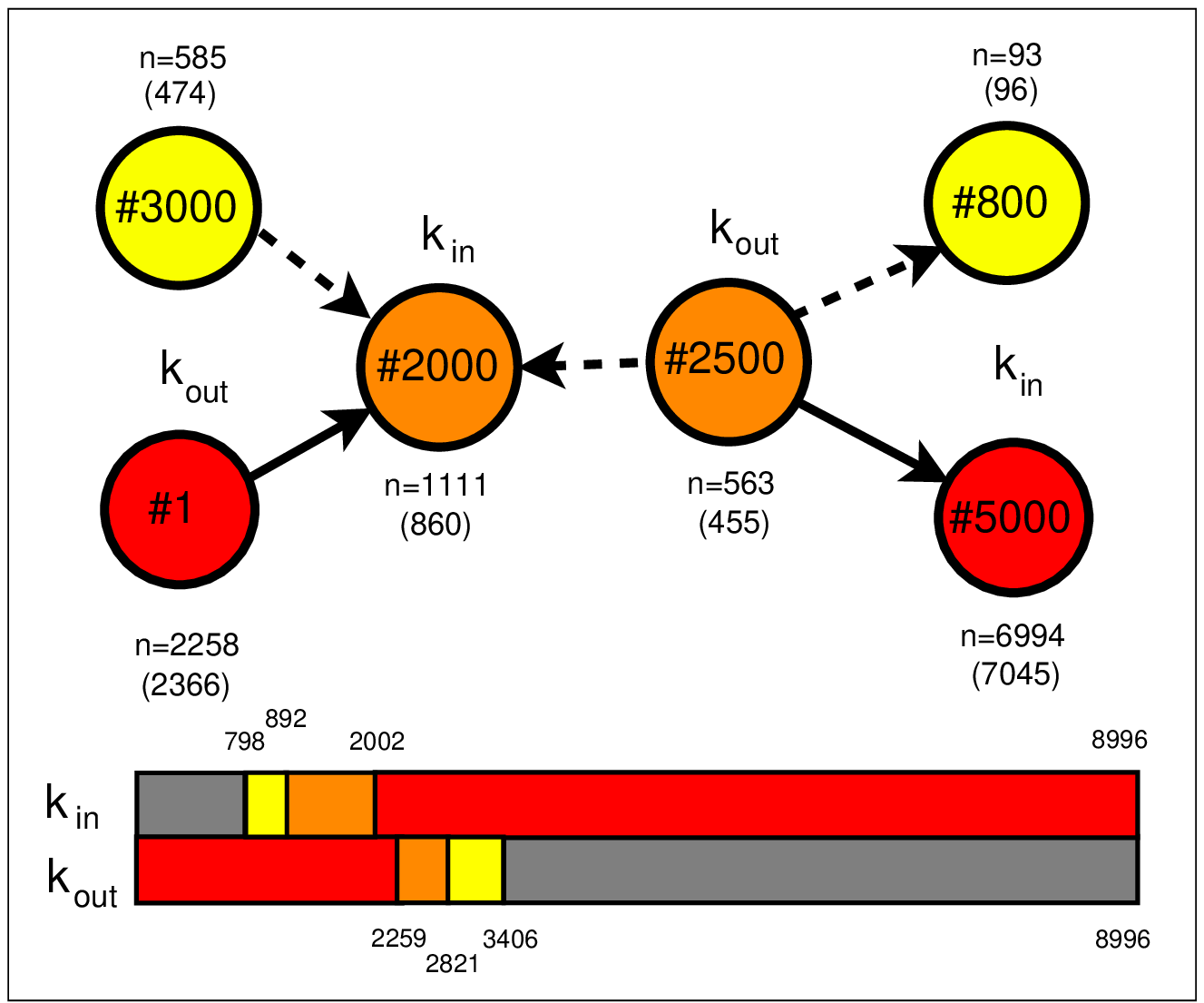}
\end{center}
Fig-2.
\begin{center}
\includegraphics[width=10cm]{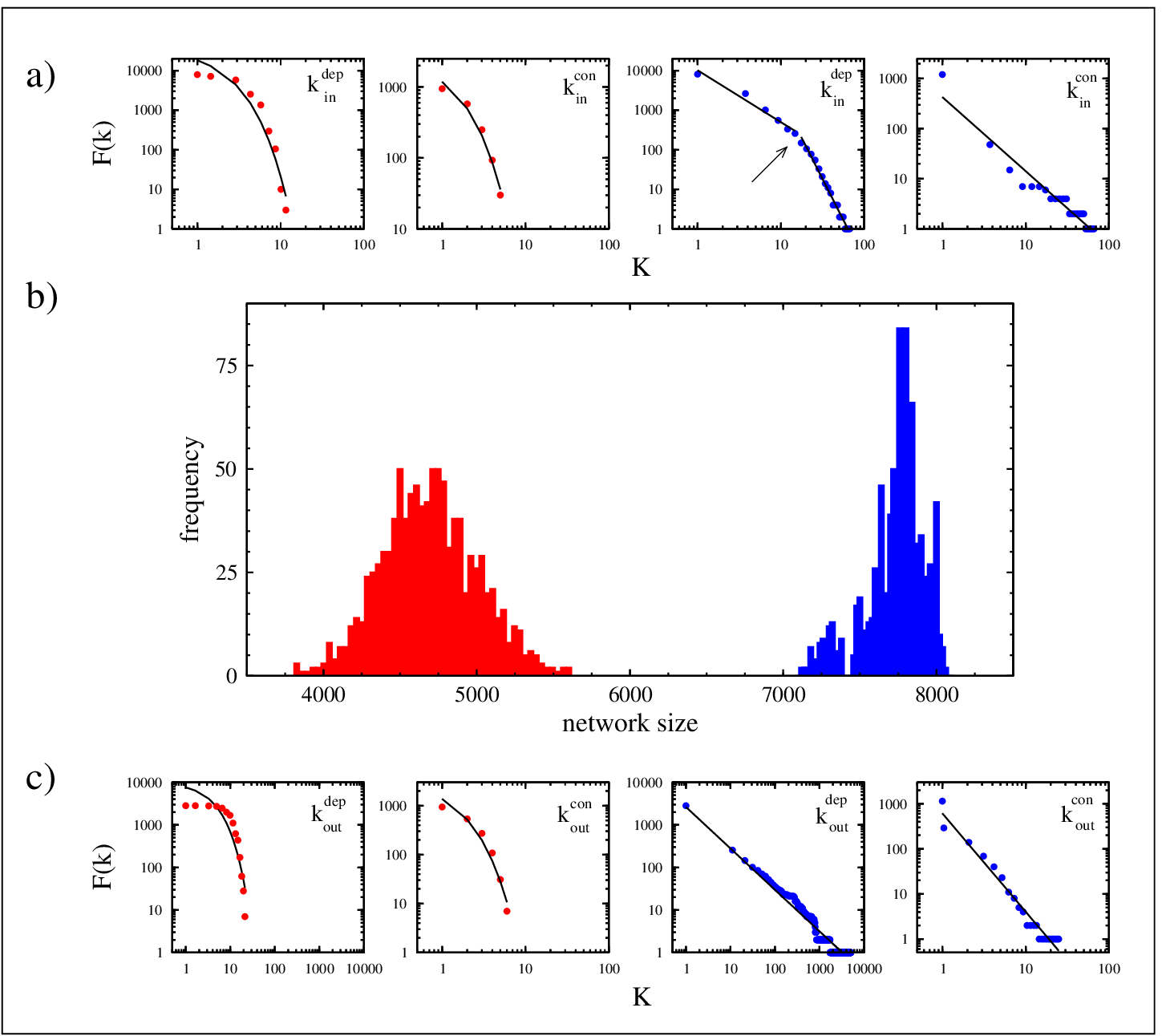}
\end{center}
\newpage
\section*{Figure Legend}
Fig. 1\\
Hypothetical graph illustrating the type of packages as a function
of their $k_{in}$ (number of incoming edges per node) and $k_{out}$
(number of outgoing edges per node) and types of interactions (solid
arrows represent dependences $k^{dep}$(number of dependences per
node), and dotted arrows conflicts $k^{con}$(number of conflicts
per node)). 
Packages with $k_{in}^{dep}$ $>0$ (e.g., package number $5000$),
$k_{in}^{con}$ $>0$ (e.g., package number $800$) or both (e.g., package
number $2000$), mean that they depend or have
a conflict with other packages, or both, respectively. 
Packages with $k_{out}^{dep}$ $>0$
(e.g., package number $1$) or $k_{out}^{con}$ $>0$ (e.g., package number
$3000$) or both (e.g., package number $2500$), mean that other packages
depend or enter into conflict with them, or both, respectively. 
Total number of packages with each type of incoming and outgoing link 
in the network is $n$
(in brackets the average value after $1,000$ replicates
of the null model; see step (2) of the null model). 
Colours in the horizontal bars correspond to the
number of each type of package in the null model. 
Yellow are packages with $k_{in}^{con}$ $>0$ or $k_{out}^{con}$ $>0$. 
Red are packages with $k_{in}^{dep}$ $>0$ and/or $k_{out}^{dep}$ $>0$. 
Orange are packages with $k_{in}^{dep}$ $>0$ and $k_{in}^{con}$ $>0$,
or $k_{out}^{dep}$ $>0$ and $k_{out}^{con}$ $>0$. 
Gray regions are packages with $k_{in}$ $=0$ or $k_{out}$ $=0$ interactions
(not shown in the graph).

\newpage
Fig. 2\\
a) Cumulative $k_{in}$ degree distributions of null model (red circles) 
and real data (blue circles). All degree distributions are marginally 
significant for both null model ($k_{in}^{dep}$, $n$=$7894$; $k_{in}^{con}$, $n$=$944$), 
and real data ($k_{in}^{dep}$, $n$=$8105$; $k_{in}^{con}$, $n$=$1204$), decaying 
exponentially ($P=0.07$, and $P=0.07$ respectively) for the null model, and 
as a power law for real data ($P=0.1$ for the first regression, and $P=0.1$ 
for the second with a breakpoint in $k=15$ (solid arrow), and $P=0.07$ respectively). 
Degree distribution of the null model represents the average value for ten replicates.

b) The size frequency distribution differs from a normal distribution
for real data (blue, Jarque-Bera test, $P<0.05$, with an average
network size of $7,647$ packages) and does not differ from a normal
distribution for the null model (red, Jarque-Bera test $P=0.2$, with
an average network size of $4,750$ packages). No replicate from the null 
model distribution is equal or higher than any replicate from the real data 
distribution ($P<0.0001$).

c) Cumulative $k_{out}$ degree distributions of null model (red circles) 
and real data (blue circles). Degree distributions for the null model are 
significant ($k_{out}^{dep}$, $n$=$2821$), and marginally 
significant ($k_{out}^{con}$, $n$=$941$), decaying exponentially in both 
cases ($P<0.05$ and $P=0.09$ respectively). Degree distribution for real 
data are significant ($k_{out}^{dep}$, $n$=$2821$), and marginally 
significant ($k_{out}^{dep}$, $n$=$1148$), decaying in both cases as a power 
law ($P<0.05$ and $P=0.08$ respectively). Degree distribution of the null 
model represents the average value for ten replicates.

\end{document}